# Reactive optical matter: light-induced motility in electrodynamically asymmetric nano-scale scatterers

Yuval Yifat,[1,†] Delphine Coursault,[1,†] Curtis W. Peterson,[1,2,†] John Parker,[1,3,†] Ying Bao,[1,4] Stephen K. Gray,[5] Stuart A. Rice,[1,2,] and Norbert F. Scherer[*,1,2]

1. James Franck Institute, The University of Chicago, 929 E. 57th Street, Chicago, Illinois 60637, USA

2. Department of Chemistry, The University of Chicago, 929 East 57th Street, Chicago, Illinois 60637, USA

3. Department of Physics, The University of Chicago, 929 East 57th Street, Chicago, Illinois 60637, USA

4. Present address: Department of Chemistry, Western Washington University, 516 High Street, Bellingham, WA 98225, USA

5. Center for Nanoscale Materials, Argonne National Laboratory, 9700 South Cass Avenue, Argonne, Illinois 60439, USA

†These authors contributed equally

*Corresponding Author, e-mail: nfschere@uchicago.edu

**From Newton's third law, the principle of actio et reactio[1], we expect the forces between interacting particles to be equal and opposite. However, non-reciprocal forces can arise.[2] Specifically, this has recently been shown theoretically in the interaction between dissimilar optically trapped particles mediated by an external field.[3] As a result, despite the incident external field not having a transverse component of momentum, the particle pair experiences a force in a direction transverse to the light propagation direction.[3,4] In this letter, we directly measure the net non-reciprocal force in electrodynamically interacting nanoparticle dimers illuminated by plane waves and confined to pseudo one-dimensional geometries. We show by electrodynamic theory and simulations that inter-particle interactions cause asymmetric scattering from heterodimers and therefore, the non-reciprocal forces are a consequence of momentum conservation. Finally, we demonstrate experimentally that non-reciprocal dynamics occur generally for illuminated asymmetric scatterers.**

There is tremendous interest and effort in the development of light-driven nanomotors, devices that convert light energy into autonomous motion.[5] Various optical methods can produce rotational motion[6] or, using primarily photo-reactive materials, translational motion.[7] A promising direction towards such nanomotors has risen from recent theoretical work predicting that dissimilar particles trapped in a plane wave would experience a non-reciprocal net force resulting in transverse motion of a particle pair due to its asymmetric scattering. This autonomous motion is seemingly in the absence of an external driving force[3,4] in the transverse plane. Simulations showed that these non-reciprocal forces vary with interparticle separation. To date, however, there has not been an experimental demonstration of this phenomenon. In this letter we experimentally demonstrate this phenomenon, thus rectifying this deficiency. We demonstrate optical self-motility beyond particle pairs by generating and measuring translational motion of asymmetrical nanoparticle assemblies.

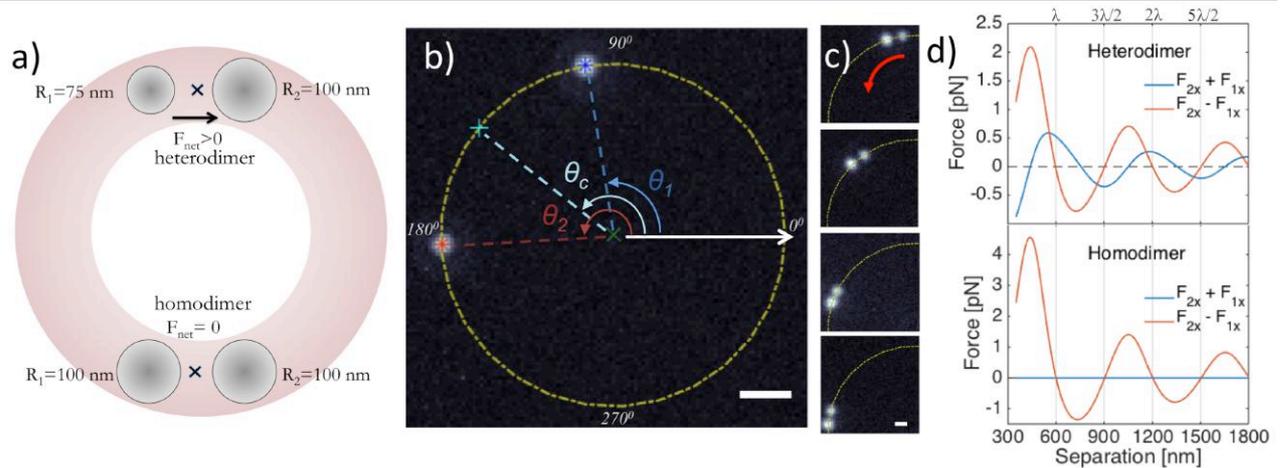

**Figure 1:** Experimental scheme for measuring non-reciprocal forces. (a) General schematic of the experiment: two dissimilar particles in a ring trap (top) experience a net force, $\vec{F}_{net} \neq 0$, resulting in observable motion. Two identical particles experience $\vec{F}_{net} = 0$ (bottom). b) Experimental image and coordinate system. Trap location is indicated by dashed yellow circle. The particle locations in the trap are $\theta_1$ and $\theta_2$. Their mean angular position is $\theta_c$. Scale bar is 1μm. (c) Directed motion event of heterodimer. When a 150 nm and a 200 nm diameter Ag NPs are at optical binding distance, we observe directed motion towards the larger particle. Time difference between frames is 75 ms, scale bar is 500 nm. (d) Sum and difference of the forces on both particles (calculated using GMT) as a function of separation for a heterodimer (top) and a homodimer (bottom). Particle sizes and orientation are identical to panel (a).

Our experiments were performed using a standard optical trapping setup with a Ti:Sapphire laser operating at $\lambda$=790nm[8,9] (see Supporting Information, SI). We used a tightly focused circularly polarized spatially phase modulated beam of light to form an optical ring trap[8,10]. A schematic of the system is shown in Fig. 1(a). We trapped a mixture of 150 nm and 200 nm diameter Ag nanoparticles (NPs) and measured their motion by dark-field microscopy at a high (290fps) frame rate. The particles positions were tracked[11–13] and their precisely determined positions used to calculate the angular position $\theta_i$ of particles $i=1,2$ on the ring. The central angle of the pair, $\theta_c$, was defined as the mean angular position of the particles (Fig. 1(b)). The particle radii were differentiated by their scattering intensity (and size) on the detector (see SI). We observed directed motion of the electrodynamically interacting pairs of dissimilar particles, termed a "heterodimer", towards the larger particle (Fig. 1c, videos S2, S3). By contrast, when two particles of the same size come into close proximity creating a "homodimer" they do not exhibit directed motion. These observations are in agreement with forces we calculated using generalized Mie theory (GMT) shown in Fig. 1d (see SI). For a stable optically bound pair[14–16] (i.e. particles separated by approximately $\lambda/n_b=600nm$ in water, where $n_b$ is the refractive index, and $\vec{F}_2 - \vec{F}_1 = 0$ ), the transverse force on the pair, $\vec{F}_{net} = \vec{F}_2 + \vec{F}_1 = 0$ only when the two particles have identical radii.[3,4]

Fig. 2(a) shows representative time trajectories of $\theta_c$ for the homodimer and heterodimers shown in the insets (videos 1-3, more trajectories shown in SI). The motion of the pair is directed toward the larger particle and can therefore move clockwise or counterclockwise around the ring depending on the heterodimer orientation. We note that the motion of the heterodimer that we observe cannot arise from solely asymmetric hydrodynamic interactions. Without an external source of momentum, hydrodynamic interaction between particles does not alter the centering of the distribution of Brownian motion displacements of each of the partners in the heterodimer around zero displacement.

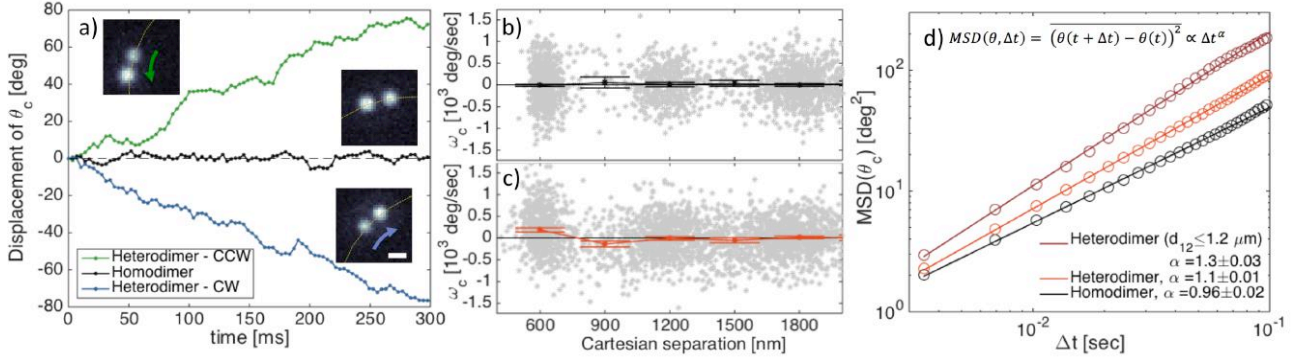

**Figure 2**: Non-reciprocal force-induced dynamics. (a) Example trajectories for homodimer (black) and heterodimer (color) moving in a counter-clockwise (green) or clockwise (blue) direction. Distribution of instantaneous angular velocities (grey dots) and the mean angular velocity of homodimers (b, black) and heterodimers (c, orange) as a function of interparticle separation. Bin size is 300 nm. Mean angular velocity value calculated by fitting a Gaussian function to the instantaneous velocity distribution. Error bars are the $3\sigma$ confidence interval for the center of the fit. Positive velocity defined as motion of the heterodimer toward the larger NP. (d) Calculated Mean Square Displacement (MSD) values for the homodimer data shown in (b) (black), heterodimer data shown in (c) (orange), and the subset of the heterodimer data where interparticle separation was $\leq 1.2\mu m$ (red). Exponents were calculated from a linear fit of the MSDs shown, individual trajectories shown in Supporting Information. Error bars are $3\sigma$ confidence intervals.

We repeated the experiment several times on different homodimers and heterodimers (see methods section and SI for full details) and combined the results. Figures 2(b,c) show the angular velocity distributions and the mean angular velocities of the dimer center, $\omega_c$, as a function of interparticle separation for the full homodimer and a heterodimer dataset, respectively. The instantaneous angular velocity $\omega_{c,n}$ is defined as the difference in the central angle of the pair in the sequential frames $n,n+1$ (i.e. $\omega_{c,n} = (\theta_{n+1} - \theta_n)/\Delta t$, $n$ is the frame number, $\Delta t$ is the time step). In an overdamped system $\omega_c \propto \vec{F}_{net}$. In order to combine data with different heterodimer orientations, we define positive velocity as the vector from the smaller particle towards the larger particle. Heterodimers exhibit a positive mean angular velocity when the particles are at optical binding separation (600±150 nm), and a negative mean angular velocity when the separation is $3\lambda/2n_b$ (i.e. 900±150 nm). By contrast, the mean angular velocity for a homodimer is zero for all separations. These observations are in accordance with our prediction from GMT calculations (see Fig. 1(d)). The change in the sign of the mean velocity between the optical binding and the $3\lambda/2n_b$ separations, and the motion of the pair towards the larger, hotter particle, suggests that the driven motion is a result of the electromagnetic field, and not a scalar property such as heating-induced self-thermophoresis[17] (see SI for full details).

Fig. 2(d) shows the average mean square displacement (MSD) of $\theta_c$ for the homo- and heterodimer trajectories. The exponent, α, of $MSD(\Delta t) = D \cdot \Delta t^\alpha$ (Diffusion coefficient D, lag time Δt) for the homodimer is α = 0.96±0.02 as expected for a diffusing Brownian particle.[18] For heterodimers we observe $\alpha > 1$, indicating driven motion[19] and of even greater value, α = 1.3 ± 0.03, when only considering trajectories when the particle separation is less than 1.2μm (two optical binding separations; a value chosen to allow longer trajectories for analysis; see SI for more details about the number of experiments and trajectories analyzed).

Recent publications calculated the dynamics resulting from an asymmetry in the linear or angular momentum of the light scattered by optically trapped objects.[20,21] We extended previous theoretical work, which treated particles in a linearly polarized beam,[3] to circular polarization to explain the self-motility of electromagnetically interacting dimers (see SI). We simulated Ag NP dimers using generalized Mie theory (GMT).[22,23] Each dimer, consisting of two spherical Ag NPs with radii $R_1$ and $R_2$ separated by a distance *d* along the *x*-axis, is placed in a water medium ($n_b = 1.33$) with an incident right-handed-circularly (RHC) polarized plane wave (of wavelength 800 nm in vacuum). Simulations were performed varying $R_2$ for three values of $R_1$ at a separation of *d* = 600 nm (Figure 3a). When $R_1 = R_2$, $\vec{F}_{net,x} = 0$ vanishes as expected for the homodimer. When $R_1 > R_2$, $\vec{F}_{net,x} > 0$, causing the heterodimer to move in the $+x$ direction. If $R_1 < R_2$, the net force is reversed and the heterodimer moves in the $-x$ direction.

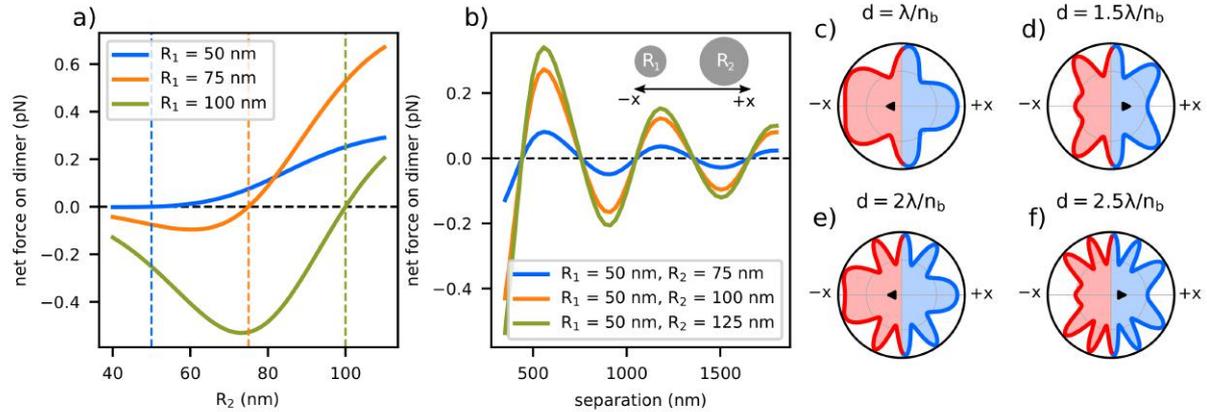

**Figure 3:** Simulations of different heterodimers using the Generalized Mie Theory (GMT) method for the force calculation. (a) Net force on the dimer, $F_{net,x}$, as a function of the radius of particle 2 with three different radius values for particle 1: 50 nm, 75 nm, and 100 nm. Dashed lines indicate the case of the three different homodimers, where $F_{net,x}$ vanishes. (b) $F_{net,x}$ vs. separation for three different heterodimers. (c-f) Angular scattering intensity in the $xy$-plane from the $R_1 = 75$nm, $R_2 = 100$nm heterodimer for different dimer separations $d$. The black triangle indicates the center of mass ('CM') of the angular distribution. We define the positive $x$ direction to be pointing from the smaller particle to the larger particle. Stable optical binding configurations ($d = \lambda, 2\lambda$) scatter more in the negative $x$ direction while unstable configurations ($d = 1.5\lambda, 2.5\lambda$) scatter more in the positive $x$ direction.

Additional simulations were performed for fixed nanoparticle radii while varying the separation, $d=\lambda/n_b, 2\lambda/n_b$. Figure 3b shows the net force on the heterodimers as a function of $d$. In each case $\vec{F}_{net,x} > 0$ at separations near 600 nm and 1200 nm, i.e. at stable optical binding configurations. For particle separations near 900 nm and 1500 nm, $\vec{F}_{net,x} < 0$, and the heterodimer is in an unstable configuration. Increasing the size of the larger nanoparticle increases $F_{net,x}$, but does not otherwise change the functional form of the force curves.

For our total system (particle and fields) to conserve linear momentum, the total momentum carried by the electromagnetic field scattered from the particle pair must be equal and opposite to the induced momentum of the dimer. Figure 3(c-f) shows a separation-dependent imbalance of angular scattering due to dipolar interference, i.e. more light is scattered in one direction than in the other. For $d = \lambda/n_b, 2\lambda/n_b$ (stable optical binding configurations), more light is scattered in the $-x$ direction and the net force acting on the dimer is in the $+x$ direction. Similarly, for $d = 3\lambda/2n_b, 5\lambda/2n_b$ (unstable configurations, see Figure 1d), more light is scattered in the $+x$ direction corresponding to a net force in the $-x$ direction. This asymmetry in the far-field angular scattering creates (non-reciprocal) forces on the dimer setting it in motion. The simulation results also confirm the switching of signs in the forces observed in experiment (Figure 2b) for different particle separations. Similar asymmetric scattering was reported for immobile plasmonic Yagi-Uda nanoantennas.[24,25]

Since the electrodynamically interacting NPs can be treated as a single (a)symmetric scatterer, a similar reactive optical matter effect, i.e. a "photophoretic" drift force,[26] is expected for particles with asymmetric shapes and asymmetric scattering[27]. We used the same experimental approach to study asymmetric NPs and aggregates; specifically, touching gold nanostar dimers and a large asymmetric aggregate of gold nanoparticles. The latter also interacts with a large number of single Au NPs in a ring trap. We use linearly polarized light instead of circularly polarized light to avoid causing the asymmetrical "particles" to rotate (spin).[28,29]

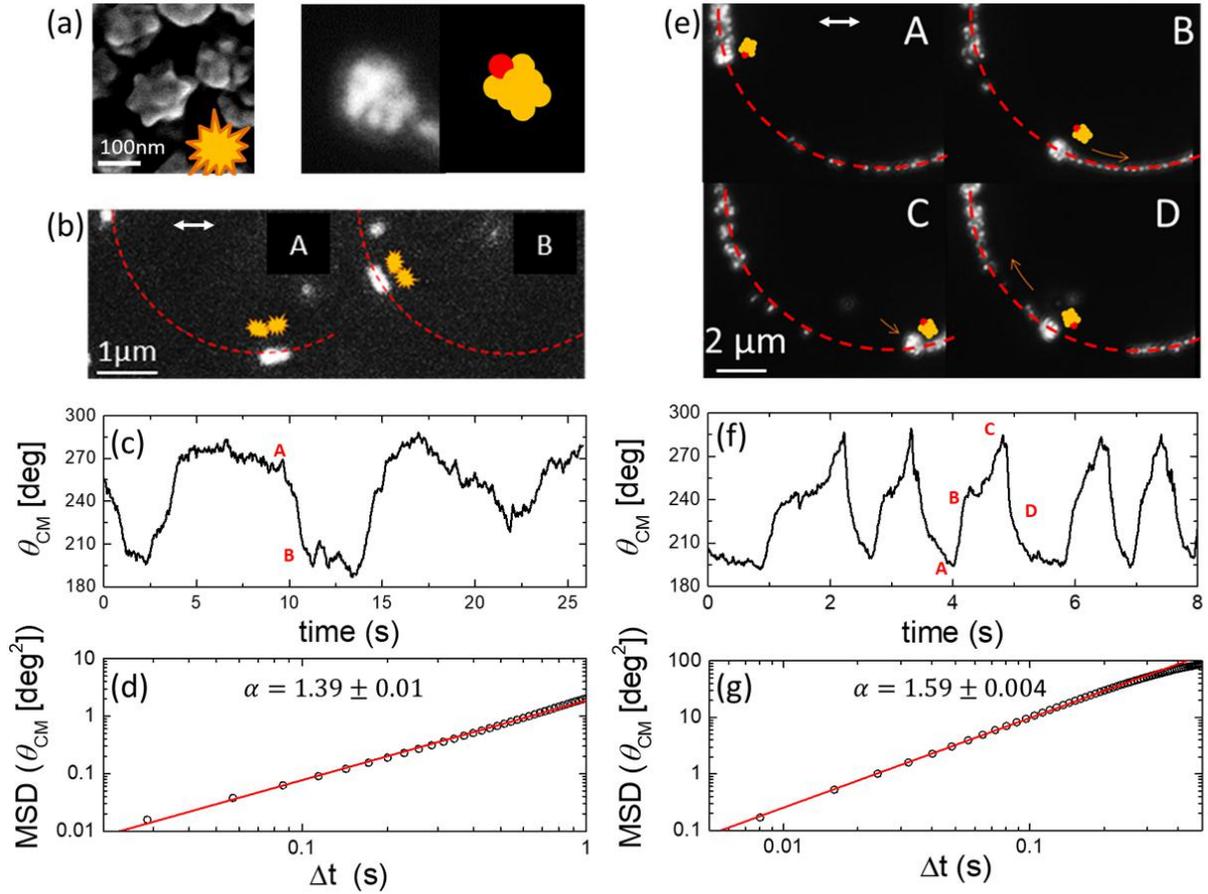

**Figure 4:** Demonstration of asymmetric forces on asymmetric nanoparticle structures. (a) SEM image of nanostars and dark-field optical image of the Au NP aggregate. Schematics (avatars) highlight their asymmetric structures and orientations. Red point defines the orientation. Dark-field images of (b) the nanostar dimer and (e) the Au NP aggregate in the ring trap (highlighted in red). The white arrow designates the polarization direction of the trapping beam. (c) and (f) show trajectories of the two asymmetrical objects. Both asymmetrical objects exhibit highly oscillatory velocity. The notable difference in their dynamics is discussed in the Supplementary Information. (d) and (g) are the MSDs calculated from the trajectories. The lower frame rate (35 fps for the nanostar, 82 fps for the aggregate) is adequate for capturing the highly driven nature of the dynamics: α = 1.39 for the nanostars and α = 1.59 for the Au NPs aggregate including the several velocity reversals.

As shown by the time-trajectory in Figure 4c, the nanostar dimer oscillates between a position parallel ($\theta_c \approx 270^o$) and perpendicular ($\theta_c \approx 180^o$) to the light polarization. It drifts tangentially to the ring and changes orientation at the $180^o$ and $270^o$ extremes of its range of motion. The restricted range of motion and switching of orientation results from its interaction with the polarized light but also from occasional interactions with neighboring nanostars (video S4). The several large variations of the velocity and the resultant MSD calculated from the time-trajectory confirm the directed (driven) motion with α ≈ 1.39±0.01; *i.e.* the nanostar dimer is strongly driven in the parts of the trajectory transitioning between the $180^o$ and $270^o$ limits. See SI for further explanation; the exact mechanism and further study of the motion control is the object of ongoing research.

Similar results are obtained for the Au-NP aggregate. Figure 4e shows that it oscillates between ≈ $180^o$ and ≈ $270^o$. It exhibits driven motion with the orientation shown by the avatars in Figure 4a,e (reversing orientation at the two extreme positions; see SI). MSD analysis shows strong driven behavior with α = 1.59. See SI for further details and discussion.

In this letter we have experimentally demonstrated driven motion of both Ag NP heterodimers and intrinsically asymmetric scatterers in optical ring traps; i.e. 1-D plane wave fields. Our electrodynamic simulations indicate the net force on a dimer is accompanied by a net asymmetric scattering in the opposite direction. We therefore attribute the driven (reactive) motion of asymmetric optical matter systems to conservation of linear momentum. While these experiments were confined to a ring trap, the results are generally applicable to any optically trapped matter structure exhibiting an electromagnetic asymmetry.

Generating directed motion at the nanoscale is challenging[30] due to the over-damped nature of dynamics at low Reynolds number and the Brownian forces that are antithetical to orientational control of nanoscale objects. Optical trapping offers a variety of solutions to these challenges since it allows precise control over the position and orientation of trapped particles. Although systematic driving forces can be applied through the use of phase gradients, non-reciprocal forces create self-motile particles that do not require specific chemical environments[31] or complex structures.[32] Therefore, optically controlled asymmetric nanoparticle assemblies, such as those reported here, can be used as active colloids[31] and fully controllable "nano-swimmers" for research in soft condensed matter and biophysics.

## Methods

**Optical trap details**: We used a CW Ti:Saphire laser with a wavelength of 790 nm (vacuum) to form a ring trap with a radius of 3.4 μm. The trap dimension was chosen in order to minimize the effect of scattering forces from particles in other sections of the ring. The laser beam was focused into a sample cell containing 150 nm diameter and 200 nm diameter Ag nanoparticles coated with a ligand layer of polyvinylpyrrolidone (PVP; NanoComposix) diluted in 18 MΩ deionized water at a ratio of 1:200.

**Particle imaging and tracking**: Following the data acquisition, we tracked the particle positions using the Mosaic particle tracking toolbox for ImageJ[11]. Due to the small size of the particles on the detector, we applied the localization algorithm with a small fitting window. This introduced pixel locking, in which the particle positions were localized towards the center of the pixels. The pixel locking was removed by applying the Single Pixel Interior Fill Factor (SPIFF) algorithm[12,13].

**Particle characterization**: The 150 and 200 nm diameter Ag nanoparticles were differentiated by imaging them on the sCMOS array detector (Andor, Neo) and observing differences in their relative size and brightness. The 200 nm diameter particles appeared larger on the sCMOS (*i.e.* occupied more pixels on the detector) and brighter compared to the 150 nm particles. We coupled the dark-field scattered light out through the side port of the inverted microscope and into a spectrometer (Shamrock-Andor SR 193i-BI-SIL) to measure the spectral response of individual particles and compare it to Mie theory scattering calculations to estimate the individual particle sizes. Full details are given in the Supporting Information.

**Data analysis and combination:** We performed 11 independent experiments, each of which was 7,000 frames in length. Of these experiments we limited the analysis to cases in which we observed two particles in the trap without a third particle near by. We then used the intensity information from the sCMOS detector to identify whether the particle pair was a homodimer (5 particle pairs, 8,500 frames) or a heterodimer (12 particle pairs, 18,900 frames). These combined data enabled us to bin the mean angular velocity $\omega_c$ as a function of interparticle separation, as shown in Fig 2(b,c) and calculate the MSD and the transport exponent, $\alpha$, for all ranges of interparticle separation. Separation dependent MSD curves were calculated by first identifying 9 trajectories of homodimer pairs and 11 trajectories of heterodimer pairs that were at optical binding separation (less than 1.2 µm). We then used their trajectories to calculate the red MSD curve shown in figure 2(d). Full details and time trajectories of the homodimers and heterodimers are given in the Supplementary Information.

**Author Contributions:** All authors participated in the writing of the manuscript.

**Conflict of Interest:** The authors declare no conflict of interest.

**Supporting Information:** Supporting Information about the experiment and theory in attached pdf file. Supporting videos S1-S5.

**Acknowledgements**

The authors acknowledge support from the Vannevar Bush Faculty Fellowship program sponsored by the Basic Research Office of the Assistant Secretary of Defense for Research and Engineering and funded by the Office of Naval Research through grant N00014-16-1-2502. We thank the university of Chicago research Computing Center for an award of computer time required for the GMT simulations. We thank the University of Chicago NSF-MRSEC (DMR-0820054) for central facilities support. We also thank Dr. Tian-Song Deng for his help in the characterization experiments of the Ag nanoparticles

**References:**

1. Newton, I. *Philosophiae naturalis principia mathematica.* (Jussu Societatis Regis ac Typis Josephi Streater. Prostat apud plures Bibliopolas., 1687).

2. Ivlev, A. V. *et al.* Statistical mechanics where newton's third law is broken. *Phys. Rev. X* **5,** 011035 (2015).

3. Sukhov, S., Shalin, A., Haefner, D. & Dogariu, A. Actio et reactio in optical binding. *Opt. Express* **23,** 247–252 (2015).

4. Karásek, V., Šiler, M., Brzobohatý, O. & Zemánek, P. Dynamics of an optically bound structure made of particles of unequal sizes. *Opt. Lett.* **42,** 1436 (2017).

5. Chen, H., Zhao, Q. & Du, X. Light-powered micro/nanomotors. *Micromachines* **9,** 41 (2018).

6. Shao, L. & Käll, M. Light-Driven Rotation of Plasmonic Nanomotors. *Adv. Funct. Mater.* 1706272 (2018). doi:10.1002/adfm.201706272

7. Xu, L., Mou, F., Gong, H., Luo, M. & Guan, J. Light-driven micro/nanomotors: from fundamentals to applications. *Chem. Soc. Rev.* **46,** 6905–6926 (2017).

8. Figliozzi, P. *et al.* Driven optical matter: Dynamics of electrodynamically coupled nanoparticles in an optical ring vortex. *Phys. Rev. E* **95,** 022604 (2017).

9. Sule, N., Yifat, Y., Gray, S. K. & Scherer, N. F. Rotation and Negative Torque in Electrodynamically Bound Nanoparticle Dimers. *Nano Lett.* **17,** 6548–6556 (2017).

10. Roichman, Y., Grier, D. G. & Zaslavsky, G. Anomalous collective dynamics in optically driven colloidal rings. *Phys. Rev. E* **75,** 20401 (2007).

11. Sbalzarini, I. F. & Koumoutsakos, P. Feature point tracking and trajectory analysis for video imaging in cell biology. *J. Struct. Biol.* **151,** 182–195 (2005).

12. Burov, S. *et al.* Single-pixel interior filling function approach for detecting and correcting errors in particle tracking. *Proc. Natl. Acad. Sci.* 201619104 (2016). doi:10.1073/pnas.1619104114

13. Yifat, Y., Sule, N., Lin, Y. & Scherer, N. F. Analysis and correction of errors in nanoscale particle tracking using the Single-pixel interior filling function (SPIFF) algorithm. *Sci. Rep.* **7,** 16553 (2017).

14. Burns, M. M., Fournier, J. M. & Golovchenko, J. A. Optical binding. *Phys. Rev. Lett.* **63,** 1233–1236 (1989).

15. Dholakia, K. & Zemánek, P. Colloquium: Gripped by light: Optical binding. *Rev. Mod. Phys.* **82,** 1767–1791 (2010).

16. Yan, Z. *et al.* Guiding spatial arrangements of silver nanoparticles by optical binding interactions in shaped light fields. *ACS Nano* **7,** 1790–1802 (2013).

17. Jiang, H. R., Yoshinaga, N. & Sano, M. Active motion of a Janus particle by self-thermophoresis in a defocused laser beam. *Phys. Rev. Lett.* **105,** 268302 (2010).

18. Einstein, A. On the motion of small particles suspended in liquids at rest required by the molecular-kinetic theory of heat. *Ann. Phys.* **17,** 549–560 (1905).

19. Metzler, R. & Klafter, J. The restaurant at the end of the random walk: recent developments in the description of anomalous transport by fractional dynamics. *J. Phys. A. Math. Gen.* **37,** R161 (2004).

20. Sukhov, S., Kajorndejnukul, V., Naraghi, R. R. & Dogariu, A. Dynamic consequences of optical spin--orbit interaction. *Nat. Photonics* **9,** 809 (2015).

21. Damková, J. *et al.* Enhancement of the 'tractor-beam' pulling force on an optically bound structure. *Light Sci. Appl.* **7,** 17135 (2018).

22. Xu, Y. Electromagnetic scattering by an aggregate of spheres. *Appl. Opt.* **34,** 4573–4588 (1995).

23. Ng, J., Lin, Z. F., Chan, C. T. & Sheng, P. Photonic clusters formed by dielectric microspheres: Numerical simulations. *Phys. Rev. B* **72,** 85130 (2005).

24. Li, J., Salandrino, A. & Engheta, N. Shaping light beams in the nanometer scale: A Yagi-Uda nanoantenna in the optical domain. *Phys. Rev. B* **76,** 245403 (2007).

25. Kosako, T., Kadoya, Y. & Hofmann, H. F. Directional control of light by a nano-optical Yagi--Uda antenna. *Nat. Photonics* **4,** 312–315 (2010).

26. Liaw, J. W., Chen, Y. S. & Kuo, M. K. Spinning gold nanoparticles driven by circularly polarized light. *J. Quant. Spectrosc. Radiat. Transf.* **175,** 46–53 (2016).

27. Simpson, S. H., Zemánek, P., Maragò, O. M., Jones, P. H. & Hanna, S. Optical Binding of Nanowires. *Nano Lett.* **17,** 3485–3492 (2017).

28. Tong, L., Miljković, V. D. & Käll, M. Alignment, rotation, and spinning of single plasmonic nanoparticles and nanowires using polarization dependent optical forces. *Nano Lett.* **10,** 268–273 (2010).

29. Liaw, J.-W., Chen, Y.-S. & Kuo, M.-K. Rotating Au nanorod and nanowire driven by circularly polarized light. *Opt. Express* **22,** 26005–26015 (2014).

30. Ebbens, S. J. & Howse, J. R. In pursuit of propulsion at the nanoscale. *Soft Matter* **6,** 726–738 (2010).

31. Howse, J. R. *et al.* Self-Motile Colloidal Particles: From Directed Propulsion to Random Walk. *Phys. Rev. Lett.* **99,** 48102 (2007).

32. Abendroth, J. M., Bushuyev, O. S., Weiss, P. S. & Barrett, C. J. Controlling motion at the nanoscale: rise of the molecular machines. *ACS Nano* **9,** 7746–7768 (2015).

# Supporting information for:
# Reactive optical matter: light-induced motility in electrodynamically asymmetric nano-scale scatterers

Yuval Yifat,[†,⊥] Delphine Coursault,[†,⊥] Curtis W. Peterson,[†,‡,⊥] John Parker,[†,¶,⊥] Ying Bao,[†,§] Stephen K. Gray,[∥] Stuart A. Rice,[†,‡] and Norbert F. Scherer[*,†,‡]

[†]James Franck Institute, The University of Chicago, 929 E. 57th Street, Chicago, Illinois 60637, USA

[‡]Department of Chemistry, The University of Chicago, 929 East 57th Street, Chicago, Illinois 60637, USA

[¶]Department of Physics, The University of Chicago, 929 East 57th Street, Chicago, Illinois 60637, USA

[§]Present address: Department of Chemistry, Western Washington University, 516 High Street, Bellingham, WA 98225, USA

[∥]Center for Nanoscale Materials, Argonne National Laboratory, 9700 South Cass Avenue, Argonne, Illinois 60439, USA

[⊥]These authors contributed equally

E-mail: nfschere@uchicago.edu

# Experimental setup

A diagram of the setup used to trap the 150nm and 200nm Ag nanoparticles is shown in Fig S1(a). The setup consisted of a continuous wave Ti:sapphire laser emitting linearly polarized light at a wavelength of 790 nm. The beam was collimated and reflected off a spatial light modulator (SLM; BNS/Meadowlark HSPDM512-785nm), which was used to shape the beam by imparting the phase necessary for a ring trap. The phase mask used in the experiment is shown in Fig S1(b). The beam was reflected off a dichroic mirror and into a Nikon Ti inverted optical microscope, through a quarter wave plate, which is used to control its polarization (*i.e.* convert from linear to circular), and through a 60x IR corrected water immersion objective (Nikon 60x Plan APO IR water immersion objective, NA=1.27). The total optical power of the trapping laser measured before the dichroic mirror was 150 mW, giving a power density of $1.5\,\text{MW/cm}^2$ at the focus.

Fig S1(c) is an image of the ring trap. In order to measure the beam dimensions we removed the near-IR filter before the sCMOS detector and imaged the reflection of the beam off the coverslip. The beam was focused slightly beneath the top coverslip. The ring was measured to have a radius of 3.4 $\mu$m and a 500 nm width (i.e. FWHM) of the annulus. In addition to the ring trap that was used in the experiments, there was a noticeable focused Gaussian beam in the center of the ring trap (i.e. a zero-order reflection from the SLM) that had no effect on the experimental results due to its large distance (R=3.4 µm) from the particle locations on the ring.

The beam was focused into a sample cell that was filled with a solution of 150 nm and 200 nm silver nanoparticles coated with polyvinylpyrrolidone (PVP). The stock solutions were diluted in 18 M$\Omega$ DI water at a ratio of 1:200. The particles were illuminated using a dark-field condenser, and the light they scattered was captured by the objective and imaged onto a sCMOS detector (Andor Neo; 6.5 $\mu$m pixel size) with a total magnification of 90x, giving an effective pixel size of 72 nm. The particle motion was captured in a 120x120 pixel region of interest on the detector with an exposure time of 1 ms at a frame rate of 289 frames

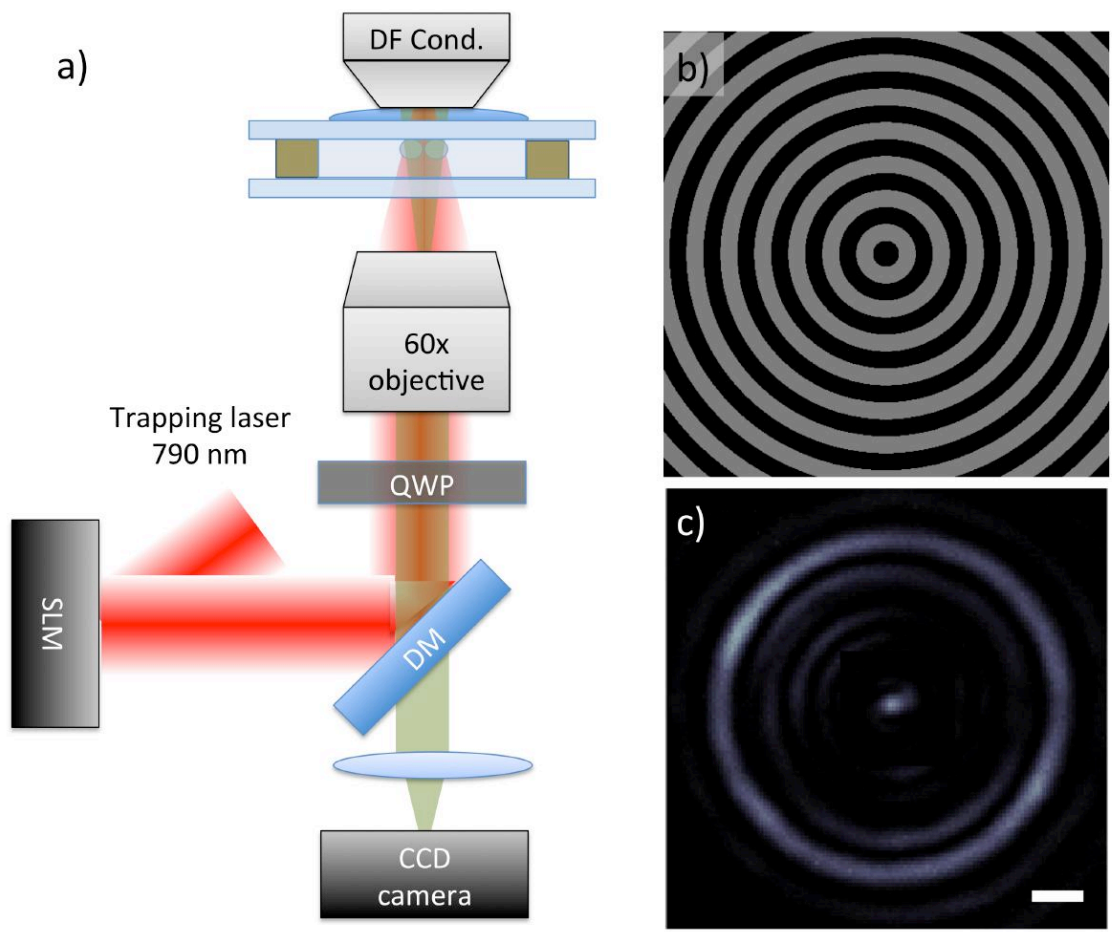

Figure S1: (a) Diagram of experimental trapping setup described in the text. SLM-Spatial Light Modulator, DF Cond. - Dark-field condenser, DM - Dichroic mirror. QWP Quarter wave plate. Trapping laser is reflected from the SLM which is used for beam shaping. Dark-field (high N.A.) illumination that scatters from the Ag nanoparticles is collected by the microscope objective, spectrally filtered and imaged to a sCMOS array detector. (b) The phase mask used to create the ring trap used in our experiments. The phase mask only uses two phase levels (black pixel level = 0 phase shift, gray pixel level = $\pi$ phase shift). (c) Image of the ring trap on the sCMOS. The Gaussian Spot in the center is the zero order reflection of the trapping laser off the SLM. The spot did not affect our experiments due to the large diameter of the trap. Scale bar is 1$\mu$m.

per second.

## Characteristics of Ag nanoparticles

The particles used for the trapping experiments described in the main text are an equal mixture of 150 nm diameter and 200 nm diameter PVP coated silver nanoparticles (NanoComposix; 150 nm diameter: NanoXact Silver KJW1882 0.02 mg/ml; 200 nm diameter: NanoXact Silver DAC1326 0.02 mg/ml). Equal parts from both stock solutions were diluted in DI water at a ratio of 1:200 and combined.

The identification of the different sized particles was achieved by analyzing their relative brightness and size on the sCMOS detector. See Figure S2(a,c) for representative images of a 150 nm (Fig S2(a)) and a 200 nm (Fig S2(c)) Ag nanoparticle imaged with our optical setup. The visual difference between the two particle images was verified as being due to their physical size by a spectroscopic measurement. Individual particles were captured in a Gaussian trap and the light scattered from them was directed through the side port of the microscope, into a spectrometer (Andor Shamrock 193 imaging spectrograph; SR 193i-B1-SIL), and detected with an EM-CD array detector (Andor Newton).

Figure S2(b,d) shows the spectral measurement for the particles imaged in figure S2(a,c) along with the expected scattering cross-section calculated from Mie theory[S1]. As can be seen in Figure S2(b), the spectral measurement from a trapped 150 nm diameter Ag nanoparticle is in agreement with the calculated Mie scattering. The abrupt decrease in signal at 750 nm is due to a near-IR filter placed after the dichroic mirror to block the reflected laser light. On the other hand, Figure S2(d) shows that the the spectral response of the 200 nm Ag nanoparticle is blue-shifted compared to the expected theoretical scattering for a 200 nm particle, and is in closer agreement with the spectrum calculated for a 175 nm Ag nanoparticle. Repeating this experiment for different particles showed a variance in the measured spectral response from the 200 nm particles, and a consistent spectral result from the 150 nm particles.

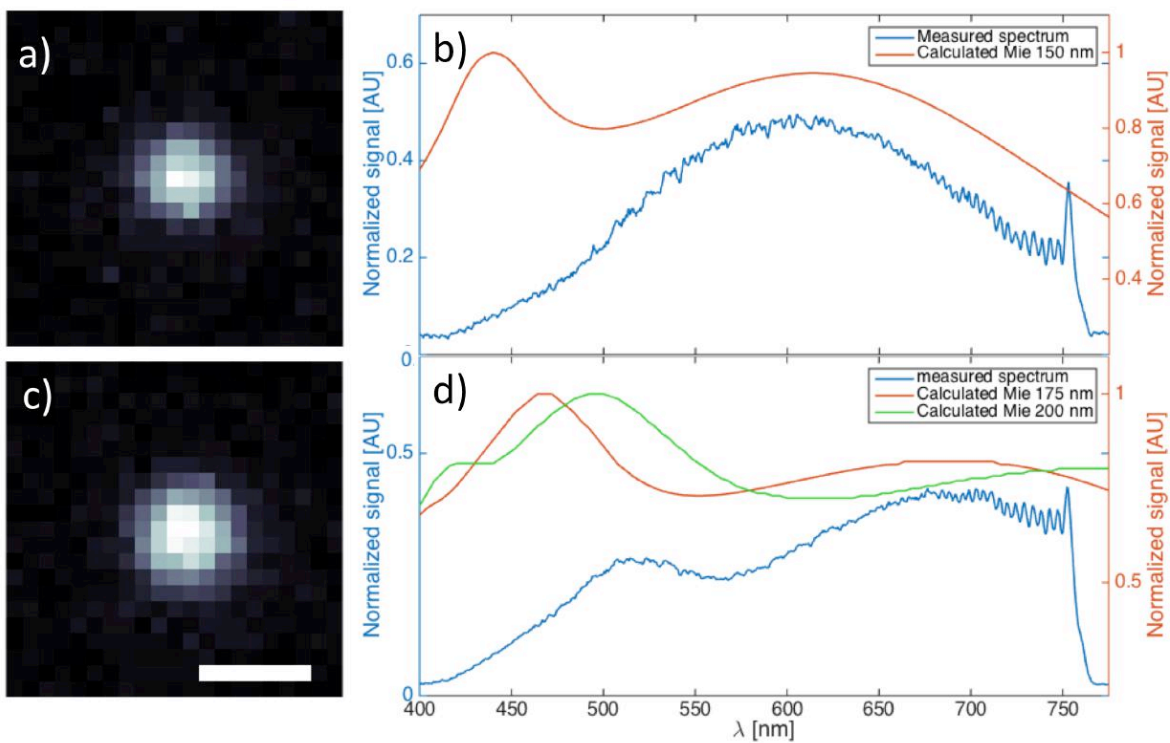

Figure S2: (a) Representative image of a 150 nm diameter Ag nanoparticle trapped in a Gaussian beam. (b) Measured scattering spectrum (blue) from the 150 nm diameter particle shown in panel (a) as well as the calculated theoretical Mie scattering for 150 nm diameter Ag nanoparticle suspended in water (red). Spectra were measured by directing the scattered light through the side port of the inverted optical microscope to a spectrometer. The abrupt drop in signal from 750 nm is due to a near-IR notch filter used to block scattered light from and reflections of the trapping beam reaching the detector. Conversely, light from 500 - 750 nm in dark-field microscopy was used to image the nanoparticles. (c) Representative image of a 200 nm diameter Ag nanoparticle trapped in a Gaussian beam. Intensity scales of images (a) and (c) are identical. Scale bar is 500 nm and applies to (a) and (c). (d) Measured scattering (blue) from the 200 nm diameter particle shown in panel (c) as well as calculated theoretical Mie scattering for 175 nm (red) and 200 nm (green) diameter Ag nanoparticles suspended in water.

The plethora of spectra for 200 nm diameter Ag nanoparticles implies dispersion in size or shape or both. This size (shape) dispersion was confirmed by electron microscopy imaging of the different sized nanoparticles. The particles were drop cast on a copper grid and imaged using Transmission Electron Microscope (TEM; FEI Tecnai F30 300kV FEG) using a magnification of x145K (see Fig S3(a,c)). The 150 nm Ag particles are uniform in size and spherical in shape, whereas the 200 nm particles were noticeably less spherical and less uniform in size.

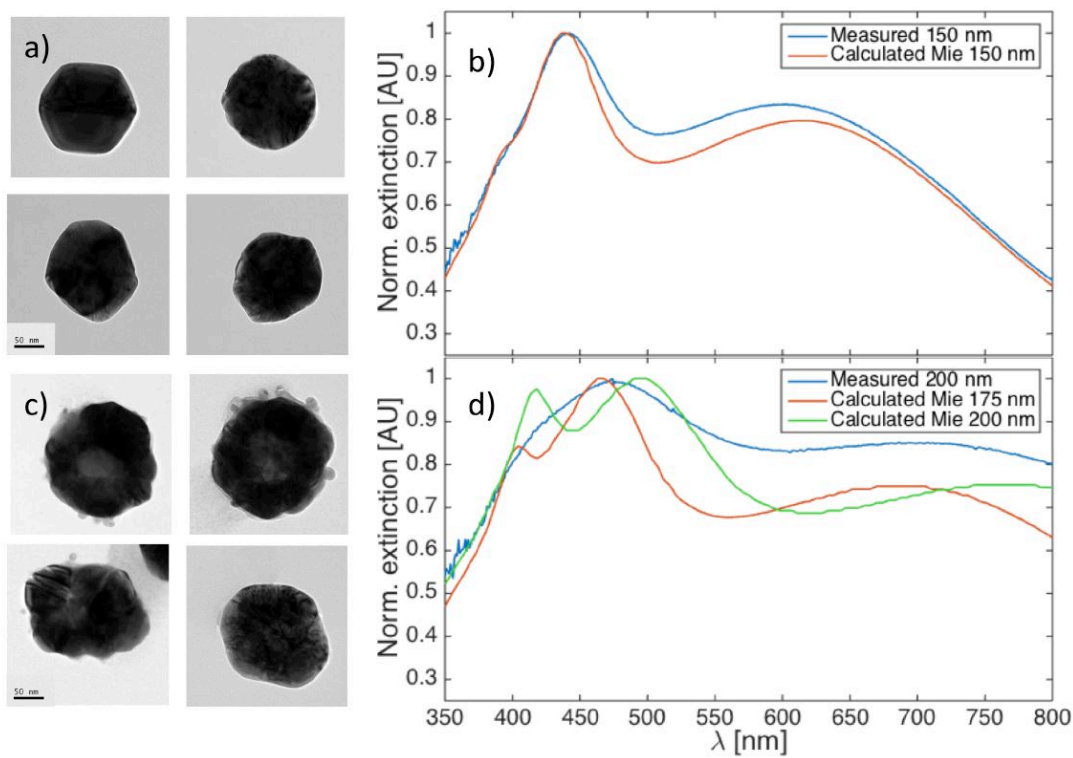

Figure S3: Transmission electron microscopy (TEM) and spectral analysis of Ag nanoparticles. (a) TEM images of 150 nm diameter Ag nanoparticles dispersed on copper grid. Scale bar is 50 nm. (b) Normalized extinction spectrum of 150nm diameter Ag nanoparticle stock solution taken with a UV-Vis-IR spectrophotometer (blue curve). Peak maxima at 440 nm and 600 nm correspond to the dipole and quadrupole Mie scattering modes of a 150 nm diameter silver particle immersed in water (calculated values given as the red curve). (c) TEM images of 200 nm diameter Ag nanoparticles dispersed on copper grid. Note the larger variance in size and shape. Scale bar is 50 nm. (d) Normalized extinction spectrum of 200nm diameter Ag nanoparticle stock solution taken with a UV-Vis-IR spectrophotometer (blue curve). Peak maxima are at 475 nm and 700 nm. Green and red curves correspond to calculated Mie extinction modes of a 175 nm and 200 nm diameter silver particle immersed in water, respectively. Note the broad width of measured peak compared to that of the the calculated values and the measured value from panel (b), implying a dispersion in particle diameters (and shapes) in the stock solution.

We also measured the ensemble extinction of the two stock solutions using a Cary 5000 UV/Vis/IR spectrophotometer (see Fig S3(b,d)). For the 150 nm diameter Ag nanoparticle solution we observed peaks at 440 nm and at 600 nm (blue curve in Fig S3(b). As there is good agreement between the measured and calculated resonance locations, these extinction peaks are the dipole and quadrupole modes calculated from Mie extinction of a silver particle of the same diameter immersed in water (red curve). However, for the 200 nm diameter Ag nanoparticle stock solution, we observed much broader extinction peaks at 475 nm and 700 nm (blue curve in Figure S3(d)). These peaks are wider than expected for a monodisperse suspension of Ag nano particles with a diameter of 175 or 200 nm (compare to red and green curves in Figure S3(d)). This implies that the solution is not monodisperse and is an ensemble of many different particle diameters with a mean value of around 180 nm. The reason for this non-uniformity results from the well-established difficulty in synthesis of Ag nanoparticles larger than 150 nm[S2].

Despite their non-uniformity, the 200 nm Ag particles are consistently larger than their 150 nm counterparts and this size difference manifests itself in the non-reciprocal dynamics shown in the main text.

# Theoretical description of non-reciprocal forces

An expression for the net optical force on a dimer (of spherical isotropic particles A and B) in the plane transverse to the propagation of plane-wave illumination can be obtained in the point dipole approximation. The component of the electric field in the $i$ direction at particles A and B is (at only one order of scattering)[S3]

$$E_A^i = E_0^i + G_{ij}^{AB} E_0^j \alpha^B; \quad E_B^i = E_0^i + G_{ij}^{BA} E_0^j \alpha^A \tag{S1}$$

where $E_0^i$ is the incident electric field, $\alpha^A$ or $\alpha^B$ is the polarizability of particle A or B, and $G_{ij}^{AB}$ are the elements of the dyadic Green's function for the vector between particles A and

B. If we assume that the particles lie on the $x$ axis, only the diagonal elements of $G_{ij}^{AB}$ are non-zero. For a circularly polarized plane wave propagating in the z direction this leads to a net force in the $x$ direction $F_x^{net}$ on the dimer

$$F_x^{net} = \frac{E_0^2}{2}\text{Re}\left[(\alpha^{A*}\alpha^B - \alpha^A\alpha^{B*})\frac{\partial}{\partial x}(G_{xx} + G_{yy}) + (\alpha^{A*}|\alpha^B|^2 - |\alpha^A|^2\alpha^{B*})\left(\frac{\partial G_{xx}}{\partial x}G_{xx}^* + \frac{\partial G_{yy}}{\partial x}G_{yy}^*\right)\right]. \quad \text{(S2)}$$

This equation extends the treatment derived in *Sukhov et. al.*[S4] from particles trapped in a linearly polarized plane wave to a plane wave with circular polarization.

The corresponding result for a pair of particles (point dipoles) interacting in a linearly polarized beam polarized along the $x$-axis (inter-particle axis) is

$$F_x^{net} = \frac{E_0^2}{2}\text{Re}\left[(\alpha^{A*}\alpha^B - \alpha^A\alpha^{B*})\frac{\partial G_{xx}}{\partial x} + (\alpha^{A*}|\alpha^B|^2 - |\alpha^A|^2\alpha^{B*})\left(\frac{\partial G_{xx}}{\partial x}G_{xx}^*\right)\right], \quad \text{(S3)}$$

which after rearrangement is identical to the result in *Sukhov et al*[S4] except for a factor accounting for infinite-order interactions between the two particles. The additional factors of $\frac{\partial G_{yy}}{\partial x}$ and $\frac{\partial G_{yy}}{\partial x}G_{yy}^*$ in equation S2 for the case of circular polarization affect the dependence of the derived forces on interparticle separation. However, equations S2 and S3 are qualitatively similar. Both equations equal zero when $\alpha^A = \alpha^B$, in accordance with the experimental and simulation results presented in the main text. In fact, both expressions vanish if the two polarizabilities are proportional by a factor of a real number (e.g. $\alpha^A = C\alpha^B$ where $C$ is a real number). *Therefore, it is necessary that the $\alpha^A$ and $\alpha^B$ have different angles in the complex plane for the non-reciprocal forces to exist within this approximation.* In summary, our analytical results show that non-reciprocal forces arise in pairs of particles with dissimilar polarizabilities under both linear and circular polarization, although the exact spatial dependence of these forces is different for those two cases.

# Analysis of combined particle trajectories

We performed 11 independent experiments, each of which was 7,000 frames in length. Of these experiments we limited our analysis to frames in which we observed two particles in the trap without another particle nearby. We then used the intensity information from the sCMOS detector to identify whether the particle pair was a homodimer (5 experimental videos, 8,500 frames) or a heterodimer (12 experimental videos, 18,900 frames).

From each video frame we localized the particle centroids using particle-tracking algorithms and used their positions to calculate the interparticle separation. The motion of their mean angular position (or "center of geometry") was calculated by how much their mean angle changed between consecutive frames, *i.e.* $\omega_n = \frac{\Delta \theta_n}{\Delta t} = \frac{\theta_{c,n+1} - \theta_{c,n}}{\Delta t}$ where $\theta_{c,n}$ is the mean angular position of the two particles in frame n, and $\Delta t$ is the time step. This data was binned by interparticle separation, $d_{12}$, and used to produce the plots of $\omega_c$ as a function of $d_{12}$ in Fig 2(b,c). By plotting the motion of the central interparticle angle as a function of interparticle separation we found the mean rotational velocity of a homodimer (Fig 2b in the main text) and a heterodimer (Fig. 2c in the main text). Fig S4 shows the distribution of $\omega_c$ along with the Gaussian fit for the homodimers and the heterodimers. Note that the FWHM of the Gaussian fits are due to the thermal Brownian fluctuations inherent in the experiment. It is important to note that the error bars shown in Figure 2(b,c) in the main text are the $3\sigma$ confidence interval for the center of the Gaussian fits, and thus, despite the width of the Gaussian distribution, its central $\omega_c$ value is statistically significant.

The MSD results and the fitted transport exponents, $\alpha$, for the entire homodimer and heterodimer dataset (*i.e.* $MSD(\theta_c | \forall d_{12})$, where $d_{12}$ is the interparticle separation) were calculated by aggregating the trajectories from the entirety of the experimental videos identified above (*i.e.* all 8,500 homodimer video frames or 18,900 heterodimer video frames). These MSD curves are shown in the main text as the black and orange curves in Fig. 2(d).

Calculation of the MSD for cases where the particles were optically bound (*i.e.* $MSD(\theta_c | d_{12} < 1.2\,\mu m))$ was done by selecting and analyzing the portions of the experimental trajectories where the interparticle separation was small enough for the particles to interact electrodynamically. Fig S5 shows the trajectories of 9 bound homodimers (a) and 12 bound heterodimers (b) from our experimental set. The start time of the trajectories was selected as when the interparticle separation was less than two optical binding separations (1.2 µm), and ended when the interparticle separation was greater than 1.5 µm for more than one time step. These values were chosen to allow analysis of cases in which the particles fluctuated away from optical binding separation for short periods of time. Fig S5(c,d) shows the calculated MSD values of the trajectories shown in Fig S5(a,b), as well as the mean MSD (connected grey diamonds) obtained by aggregating over the all the bound homodimer or heterodimer trajectories. The aggregated MSD curve for the bound heterodimer is identical to the orange curve shown in Figure 2(d) the main text. The Mean homodimer MSD shown in Fig S5(c) has a slightly different exponent than that of the entire homodimer population regardless of separation (black curve in Fig 2(d) in the main text). The reason for the slight difference between the exponent values ($\alpha = 0.96$ vs. $\alpha = 1.0$) is that the MSD shown in Fig S5(c) was fitted only on the trajectories in which the particles are at optical binding (*i.e.* $d_{12} \leq 1.2$ $\mu m$). Conversely, the exponent shown in Fig 2(d) in the main text was obtained by fitting the *entire* trajectory information. Trajectories that are shorter than 35 time steps (roughly 0.1s) are not shown in the Figure. Adding them to the aggregated data did not effect the value of $\alpha$.

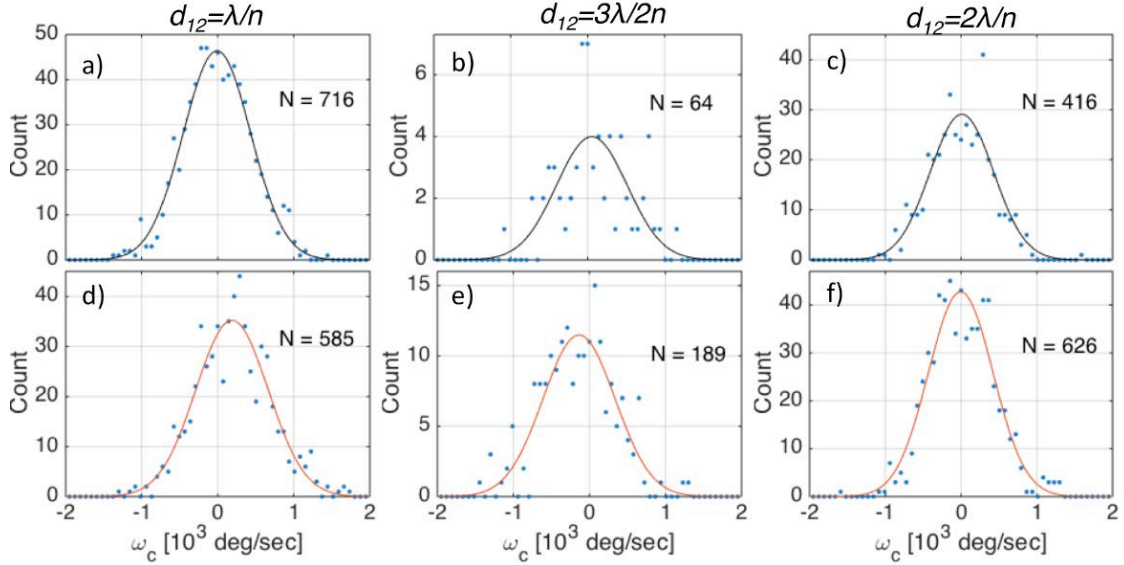

Figure S4: Distributions and Gaussian fits to the homodimer (a-c) and heterodimer (d-f) dimer velocity data shown in the main text in Figure 2(c,d). Different columns represent particles separated by one optical binding distance (a,d), 1.5 optical binding distance (b,e) and two optical binding distances (c,f). (a) Histogram of instantaneous angular velocity for homodimers where the particles are at one optical binding separation ($450 \leq d_{12} < 750$ nm). Center of the Gaussian curve is at $-9 \pm 26$ deg/sec (mean $\pm$ S.D). (b) Homodimer velocity data for the first unstable separation (($750 \leq d_{12} < 1050$ nm). Center of Gaussian fit is at $54 \pm 130$ deg/sec. (c) Homodimer velocity data for the second optical binding separation (($1050 \leq d_{12} < 1350$ nm). Gaussian center is at $11 \pm 52$ deg/sec. (d) Histogram of instantaneous angular velocity for heterodimers where the particles are at one optical binding separation ($450 \leq d_{12} < 750$ nm). Center of the Gaussian fit (orange curve) is at $190 \pm 50$ deg/sec. (e) Heterodimers velocity data for the first unstable separation (($750 \leq d_{12} < 1050$ nm). Center of Gaussian fit is at $-136 \pm 70$ deg/sec. (f) Heterodimers velocity data for the second optical binding separation (($1050 \leq d_{12} < 1350$ nm). Center of Gaussian fit is at $5 \pm 40$ deg/sec. The values of N in each panel indicate the total counts (events) in each histogram.

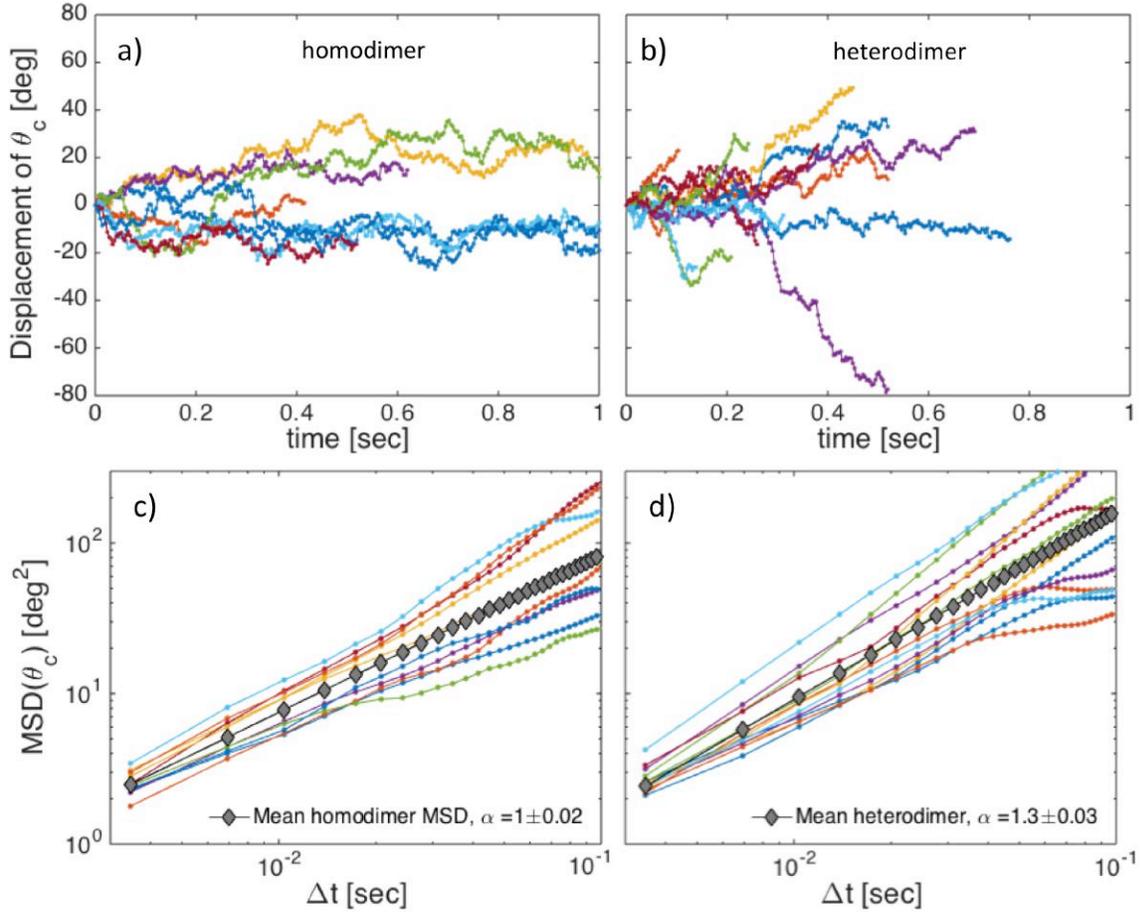

Figure S5: Experimental trajectories and MSD calculations for bound homodimer and heterodimers. (a-b) Time series of angular displacement for homodimers (a) and heterodimers (b). Trajectories were started when the particles were separated by less than 1.2 µm, and ended when the particles were separated by more than 1.5 µm for longer than one time step. This was done to include trajectories in which the particles fluctuate out of optical binding for short periods of time and to allow aggregation of long trajectories. Particle size was determined by sCMOS detector intensity data. (c-d) MSD values of trajectories for homodimer (c) and heterodimer (d). The different colors correspond to the trajectories shown in panels (a,b). The mean MSD value (marked as gray connected diamonds) is the mean MSD calculated from all the individual trajectories shown in (a,b). The mean heterodimer MSD is identical to that shown in the Figure 2(d) in the main text.

The MSD calculated from a single heterodimer trajectory demonstrates the driven nature of the heterodimer motion[S5]. Figure S6 shows an example of the MSD calculated from a single heterodimer trajectory (specifically the trajectory of the heterodimer driven in the CW direction motion shown in Fig 2(a) in the main text), along with a quadratic fit that demonstrates its driven motion.

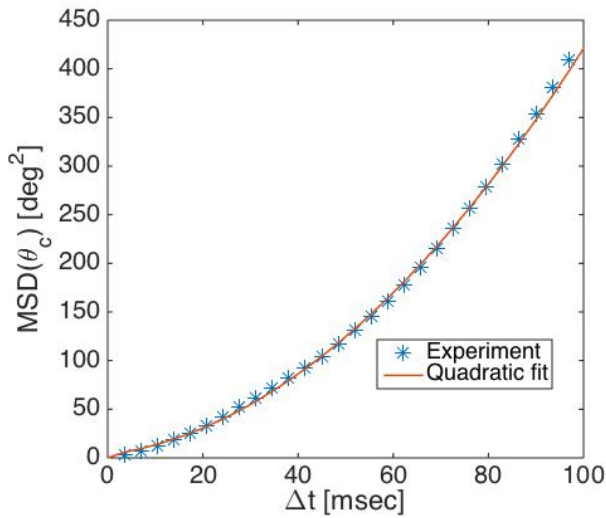

Figure S6: MSD calculated from a single heterodimer trajectory, specifically the CCW heterodimer trajectory shown in Fig 2(a) in the main text. The MSD was fitted with a quadratic function, demonstrating the driven nature of the motion.

# Effect of nanoparticle heating

It is well established that micro- and nano-scale Janus particles exhibit driven motion through self-thermophoresis.[S6,S7] A Janus particle is usually designed so that half of its surface area is coated with a material such as Au which absorbs the laser light. When such a particle is illuminated it will heat the environment around the coated area and exhibit driven motion towards its cooler, uncoated side due to increased thermal fluctuations on the heated side. The driven motion we observe in this manuscript is significantly different from self-thermophoretic motion. In this section we will describe the observed differences between self thermophoresis and electrodynamically driven motion.

The first important point: if the cause of the motion of the heterodimer were solely due to self-thermophoresis, one would expect that increasing the interparticle separation would cause a monotonic decrease in the driven component of the pair. Essentially, as the particles are further apart, they are less like a Janus particle and the driven component in their motion should decrease. However, as we show in Fig. 2c in the main text, when the particles are separated by $3\lambda/2$ we observe a statistically significant motion in the *opposite* direction (*i.e.* towards the smaller particle). This change in the direction of the directed motion cannot be explained by self-thermophoresis, and supports our observation that the motion is due to oscillatory electromagnetic interaction.

Another distinction from self-thermophoresis is the direction of the directed motion. We can calculate the heating of the particles in the trap by using the methods described in the literature[S8]. The excess temperature of the environment $\Delta T_{NP}$ around a nanosphere trapped in water near the glass coverslip is defined as $\Delta T_{NP} = \sigma_{abs} I / 4\pi R \kappa_{glass}$, where $\sigma_{abs}$ is the absorption cross section of the nanoparticles ($1.96 \times 10^3$ nm$^2$ and $3.09 \times 10^3$ nm$^2$ for the 150 nm and 200 nm diameter nanoparticles respectively), $I$ is the incident laser intensity (1.5MW/cm$^2$), $R$ is the particle radius and $\kappa$ is the thermal permittivity of glass (1.4 W/m-K), which is the dominant avenue for heat removal in our system. The result of this calculation is that the temperature difference between the particles is small ($\Delta T_{200nm} = 26.3^0$, $\Delta T_{150nm} = 22.3^0$). This slight temperature difference leads to a slight difference in dynamic viscosity of the water (0.52mPa for the 200nm nanoparticles, 0.56 mPa for the 150 nm diameter nanoparticles). If we consider only water as the sink for thermal energy from the nanoparticles ($\kappa = 0.6$ W/m-K), the particle temperature will be higher ($\Delta T_{200nm} = 61.3^0$, $\Delta T_{150nm} = 52.0^0$) and the viscosity will be lower (0.32mPa for the 200nm nanoparticles, 0.36mPa for the 150 nm diameter nanoparticles). Even in this regime, the temperature and viscosity difference between the two particles is not strikingly large.

It is not straightforward to consider our system as a Janus particle because the particles are not physically bound to each other and the separation between them changes. In addition, due to their separation (roughly 600 nm), the temperature around each individual particle will be roughly uniform. The reason for this is that the gap between the particles is significantly larger than their individual size and thus the medium directly around them will be affected, to first approximation, by the heating of the individual particles (see treatment of particle pairs in G. Baffou et al[S8].

Nevertheless, if we take the temperature difference between the particles as the cause of the driven motion, the observed motion direction is the *opposite* to what is expected for a Janus particle in water. The driven motion we have observed is towards the large particle, which is the slightly warmer particle and experiences a smaller local viscosity. In other words, based on the observed motion the Soret coefficient of our system (defined as $S_T = D_T/D$, where $D$ is the diffusion coefficient and $D_T$ is the thermodiffusion coefficient) is negative. By contrast, previous papers reported a positive Soret constant (*e.g.* motion of the Janus particle towards the cooler side) when the particle was placed in water[S7,S9,S10]. While it is possible to obtain a negative Soret coefficient by adding a surfactant to the water or by decreasing the water temperature, we performed our experiment in pure DI water and at room temperatures, and we do not anticipate any reason for a negative Soret coefficient.

Thus, the nature of the motion that we observe − the direction of the dimer motion towards the larger, hotter particle, and the dependence of the driven motion direction on interparticle separation — suggest that the driving force is not thermophoretic in nature. We conclude that the reason for the driven motion is the electrodynamic interaction between the particles, in agreement with previous theoretical work and with our simulations.

## Simulation methods

**Force evaluation through Generalized Mie Theory**

The electrodynamic interactions are computed using the Generalized Mie Theory (GMT) method.[S11,S12] In GMT, the incident and scattered fields are expanded into the vector spherical harmonic (VSH) functions for each particle. The incident field is expanded into the regular VSH's $\boldsymbol{N}_{nm}^{(1)}$ and $\boldsymbol{M}_{nm}^{(1)}$,

$$\boldsymbol{E}_{\text{inc}}^{j} = -\sum_{n=1}^{L_{\max}} \sum_{m=-n}^{n} iE_{mn} \left[ p_{mn}^{j} \boldsymbol{N}_{nm}^{(1)} + q_{mn}^{j} \boldsymbol{M}_{mn}^{(1)} \right] \tag{S4}$$

where $L_{\max}$ is the maximum number of multipole orders to expand in, $E_{mn}$ is a normalization constant, and $p_{mn}$ and $q_{mn}$ are the expansion coefficients to be solved for. The scattered field is expanded into the scattering VSH's $\boldsymbol{N}_{nm}^{(3)}$ and $\boldsymbol{M}_{nm}^{(3)}$,

$$\boldsymbol{E}_{\text{scat}}^{j} = -\sum_{n=1}^{L_{\max}} \sum_{m=-n}^{n} iE_{mn} \left[ a_{n}^{j} p_{mn}^{j} \boldsymbol{N}_{nm}^{(3)} + b_{n}^{j} q_{mn}^{j} \boldsymbol{M}_{mn}^{(3)} \right] \tag{S5}$$

where $a_n^j$ and $b_n^j$ are the ordinary Mie coefficients[S13] of particle $j$.

The expansion coefficients are solved for a system of $2NL_{\max}(L_{\max}+2)$ equations,

$$\begin{aligned} p_{mn}^{j} &= p_{mn}^{(j \to j)} - \sum_{l \neq j}^{(1,N)} \sum_{v=1}^{L_{\max}} \sum_{u=-v}^{v} A_{mn}^{uv}(l \to j) a_{v}^{l} p_{uv}^{l} + B_{mn}^{uv}(l \to j) b_{v}^{l} q_{uv}^{l} \\ q_{mn}^{j} &= q_{mn}^{(j \to j)} - \sum_{l \neq j}^{(1,N)} \sum_{v=1}^{L_{\max}} \sum_{u=-v}^{v} B_{mn}^{uv}(l \to j) a_{v}^{l} p_{uv}^{l} + A_{mn}^{uv}(l \to j) b_{v}^{l} q_{uv}^{l} \end{aligned} \tag{S6}$$

where $p_{mn}^{(j \to j)}$ and $q_{mn}^{(j \to j)}$ are the expansion coefficients of the incident source and $A_{mn}^{uv}(l \to j)$

and $A_{mn}^{uv}(l \to j)$ are VSH translation coefficients from particle $l$ to particle $j$. Solving this system includes induced dipole interactions as well as many-body interaction terms.

Once the expansion coefficients are solved for, the force on each particle can be determined by integrating the Maxwell stress tensor (MST) $\boldsymbol{T}$ over the surface of each sphere,

$$\boldsymbol{F} = \oint_\Omega \boldsymbol{T} \cdot d\boldsymbol{\Omega} \tag{S7}$$

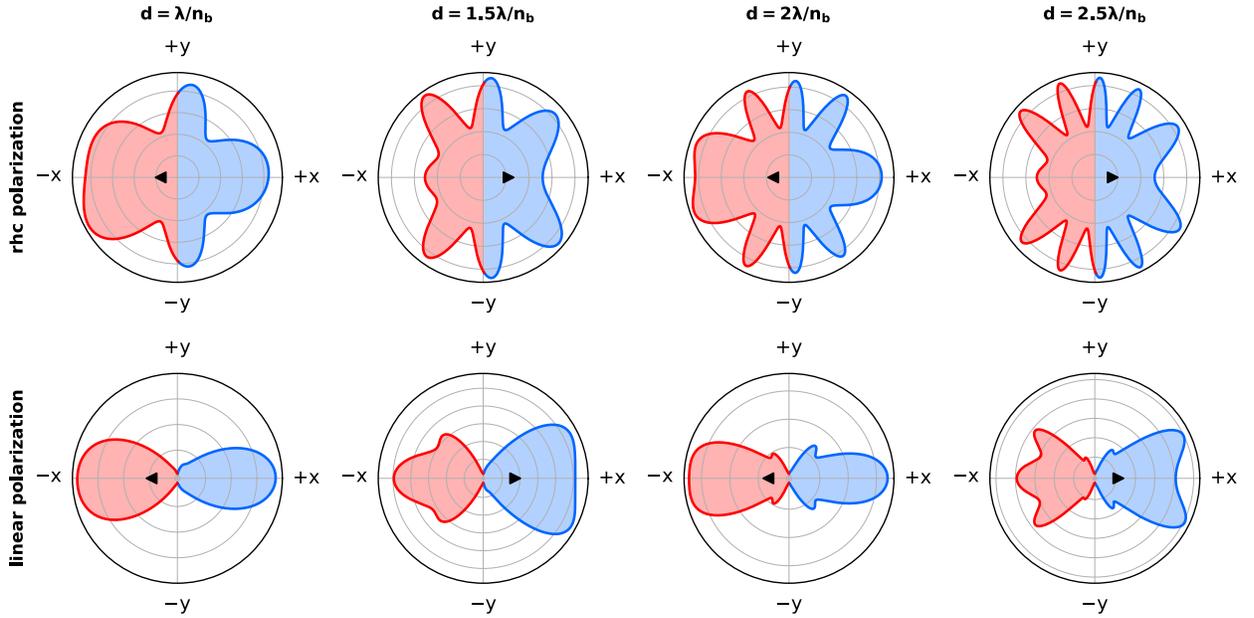

Figure S7: Angular scattering in the $xy$-plane for the hetero-dimer for different particle separations (integer and half integer multiples of the wavelength) and incident light polarization (right hand circularly polarized and linearly polarized along the y-axis). The black triangle is the centroid of the angular distribution and indicates the preferred direction of angular scattering. For separations equal to an integer multiple of the wavelength, more light is scattered in the $-x$ direction while for half integer multiples more light is scattered in the $+x$ direction.

**Langevin equation of motion**

The equation of motion for a 2-particle system undergoing dissipation and thermal noise is given by the Langevin equation

$$m_i \frac{d^2 \boldsymbol{r}_i}{dt^2} = \boldsymbol{F}_i(\boldsymbol{r}_i, t) - \lambda_i \frac{d\boldsymbol{r}_i}{dt} + \boldsymbol{\eta}_i \tag{S8}$$

S17

where $m_i$ is the mass of each nanoparticle, $\boldsymbol{F}_i$ is the electrodynamic force on each particle, $\lambda_i = 6\pi\nu R_i$ is the friction coefficient ($\nu$ is the dynamic viscosity of water), and $\boldsymbol{\eta}_i$ is a Gaussian noise term such that the fluctuation-dissipation theorem holds. Equation (S8) is integrated in time using a leap-frog Verlet integrator[S14] to give the trajectories of the nanoparticles.

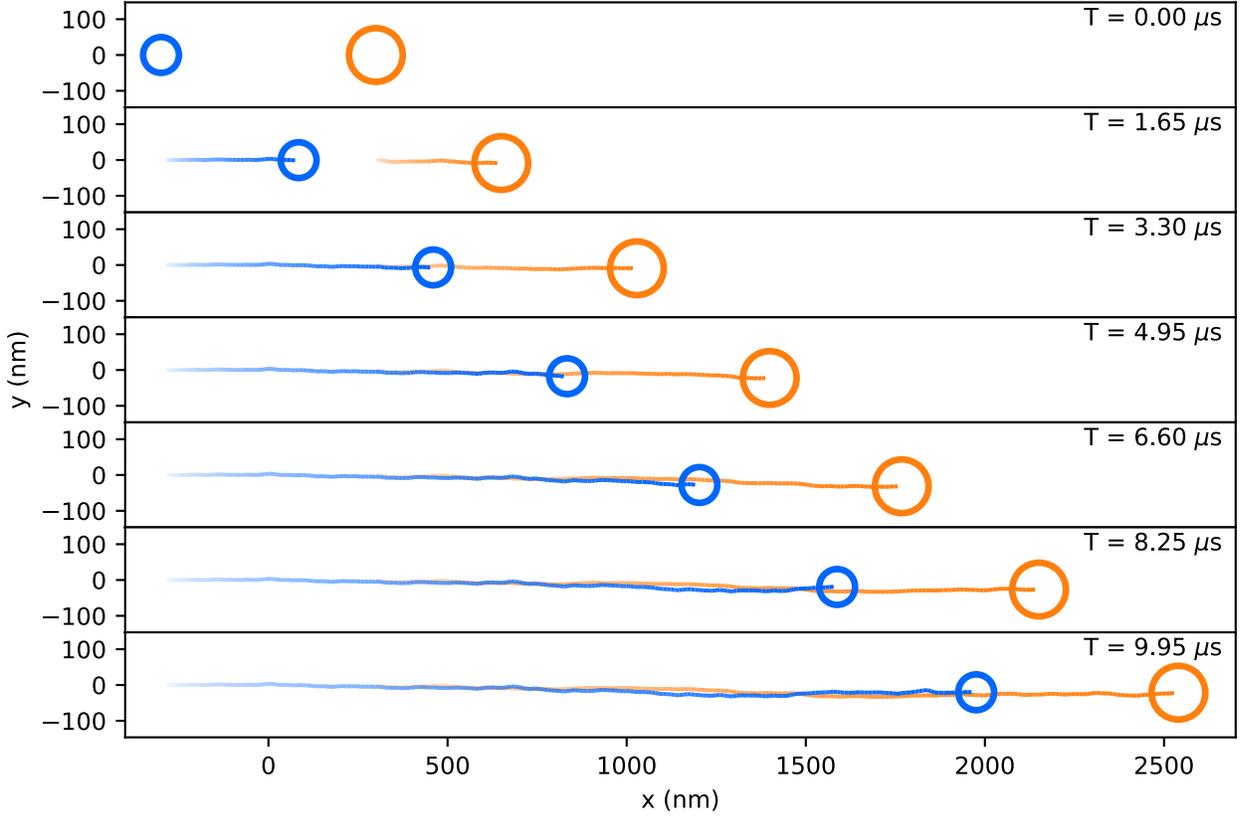

Figure S8: Trajectory snapshots of the simulated hetero-dimer using the GMT-LD method.[S15] The incident source is a y-polarized plane wave with no component of the Poynting vector in the $xy$-plane. The blue particle is 100 nm in diameter while the orange particle is 150 nm in diameter. A temperature of T = 300 K is used in a water medium (index of refraction $n_b = 1.33$). The motion of the hetero-dimer is a manifestation of the non-zero (non-reciprical) net electrodynamic force.

# Characteristics of Au nanostars

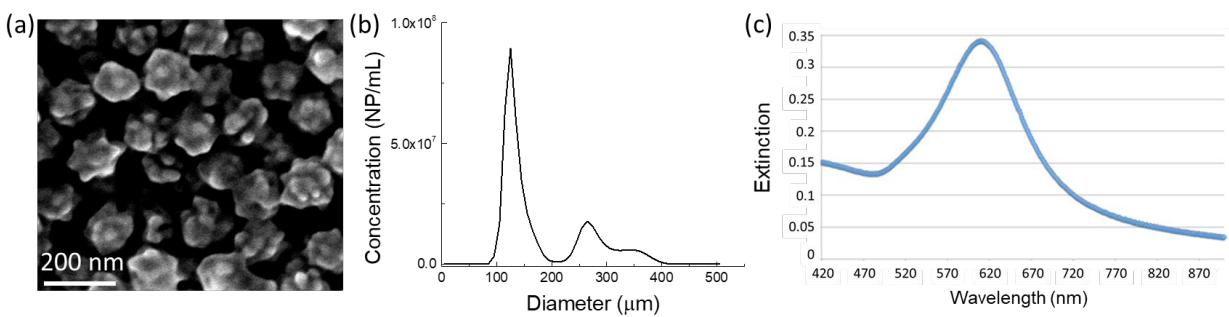

Figure S9: Characteristics of the gold nanostars. (a) Scanning electron microscopy image (SEM) images of Au nanostars. (b) Nanoparticle sizes were determined by tracking analysis (with Nanosight NS300-Malvern) reveal one major peak at 125nm diameter corresponding to the average diameter of single particles, and peaks at 265nm and 350nm. The latter reveals the significant presence of dimer and trimer aggregates in the solution. (c) Extinction spectra of the Au nanostar solution.

# Characteristics of Au nanoparticle cluster

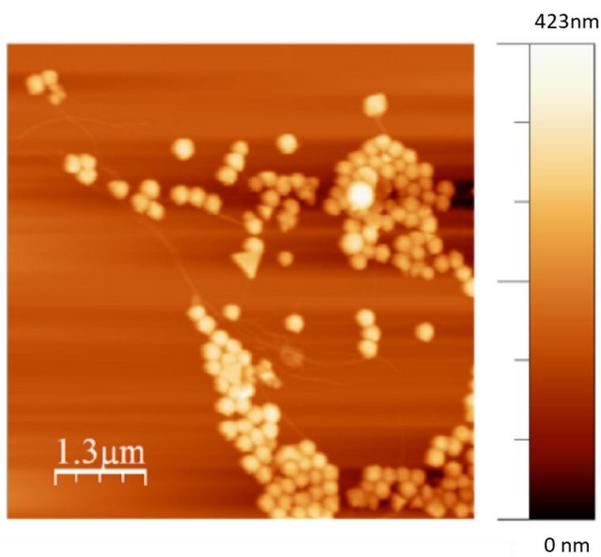

Figure S10: Characteristics of the Au nanospheres that form the large aggregate shown in Figure 4 of the main text. AFM image of the gold nanoparticles of 200nm diameter. Individual Au nanospheres as shown here are also present in the ring trap of Figure 4.

# Driven motion of Au nano-star dimer and Au NP aggregate

While circularly polarized light provides an isotropic excitation along the ring trap, we expect that the nanostar particles would spin (and the dimer would rotate except that the confinement of the ring trap would hinder rigid-body type rotation), as has been demonstrated for an anisotropic nanoparticle such as a nanorod or nanowire, or nanosphere dimer in the near field interaction regime[S16,S17]. In this manuscript we want to focus on linear driven motion and prevent the emergence of a parasitic effect such as spinning; even though understanding it will be an interesting study. Under linearly polarized light, the isotropic optical field-dimer interaction within the ring trap is broken. We expect the particles to align with the light polarization to minimize the induced torque, and to be driven due to asymmetric scattering. However, as mentioned above, the dimer cannot rotate as a rigid body, but thermal energy could cause internal rearrangements that essentially reverse the direction of the anisotropic polarizability causing a reversal in the direction of light scattering and of its motion. We observe that the dimer spends a longer time around $\Theta_c = 270^0$ (with an orientation parallel to the polarization). The nanostar dimer is able to rearrange because of the Brownian thermal noise.

There is a noticeable difference in the dynamics of the Au NP aggregate vs. that of the nanostar dimer. We believe this to result from the intrinsic scattering properties of the aggregate and the electrodynamic interactions between the aggregate and the many Au NPs present in the ring trap. In the trajectory shown from A to B in Figure 4e in the main text, the mean speed is as high as $384^o/s$ then decreases to $68^o/s$ from B to C and increases, after flipping orientation, up to $267^o/s$ between C and D. Furthermore, as shown in Figure S11, the aggregate is oriented perpendicular near $270^o$ and parallel near $180^o$ to the polarization before it flips. Several frames from the video that demonstrate this change in orientation are shown in Figure S11 for a location of the aggregate near $180^o$ in the ring.

The speed of the cluster is non-linear, it decreases near 245° which corresponds to the region where it electrodynamically "contacts" the many single Au NPs also trapped in the ring. By contrast, these interactions are negligible in the case of the nanostar dimer because of the lower particle density in the ring.

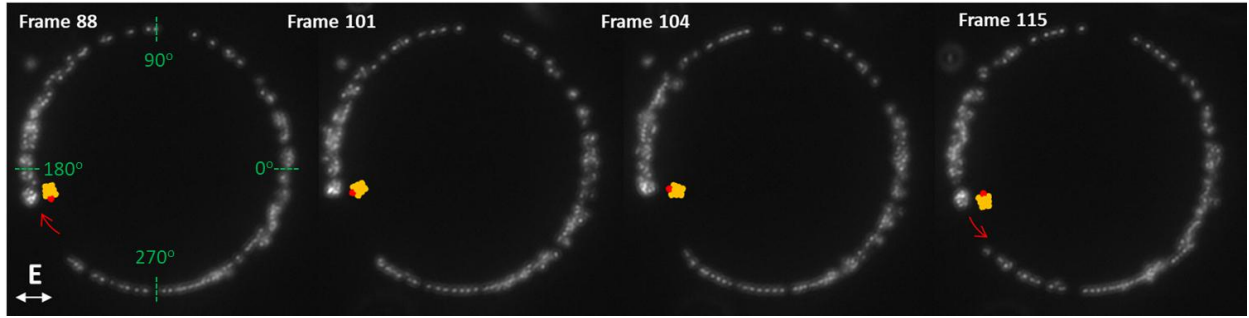

Figure S11: Dark field images of the Au NP aggregate and Au NPs optically trapped in the ring. Images are presented in chronological order showing the evolution of the cluster orientation around 200°. Frame rate is 35fps. The dense groups of individual Au nanoparticles in the $< 180^0$ and $> 270^0$ regimes are due to a slight astigmatism.

The difference in the dynamics between the dimer and the aggregate is thus not only due the intrinsic scattering properties of the aggregates but also the result of the interaction of the aggregate with the optically bound Au NPs in the ring trap. The interactions reduce the net drift force even though the behavior of the aggregate is strongly super diffusive (see Figure 4f). Conversely, the interaction and driven motion of the Au NP aggregate affect the local NP density. As observed in the video, the Au cluster pushes the Au NPs inducing a compression of Au NPs in these two regions (around 180° and 285°). Notably, the aggregate does not proceed further presumably both because of the resistance of the Au NPs to further compression and also its interaction with the linearly polarized beam.

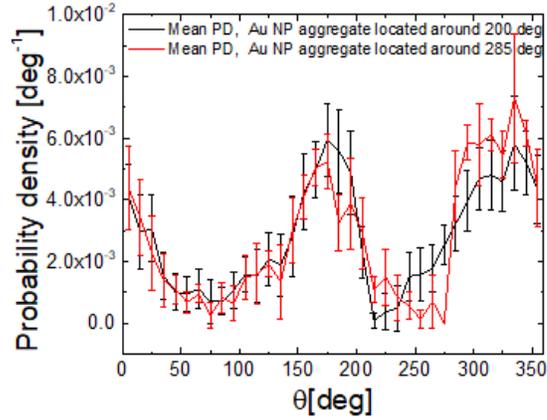

Figure S12: Evolution of the Au nanoparticle (NP) density in the ring. We show the average NP density depending on the two extreme positions of the aggregate in the ring trap. The local NP density is increased near the Au NP aggregate (i.e. near the $200^o$ and $270^o$ location). The error bars represent the standard deviation of the NP density. The Au NP density increases at angular values beyond where the Au NP aggregate goes due to its "sweeping" and then forcing them into more compact angular regions.

As shown in Figure S12, our interpretation is supported by the average probability density calculated for the extreme position of the cluster. At $200^o$ (respectively $285^o$), there is an increase (respectively a decrease) of the density of NPs around $180^o$ and a decrease (respectively an increase) of the density around $270^o$ when the Au NP aggregate is at $200^o$ ($285^o$) positions. Further investigation of the phenomenon is beyond the scope of the paper.

## List of videos

Video S1 - video of homodimer in ring trap.

Video S2 - video of heterodimer in ring trap - motion in a CW direction

Video S3 - video of heterodimer in ring trap - motion in a CCW direction

Video S4 - video of the nanostar dimer in a ring trap.

Video S5 - video of Au nanoparticle cluster in the ring trap.

# References

(S1) Horvath, H. *Journal of Quantitative Spectroscopy and Radiative Transfer* **2009**, *110*, 787–799.

(S2) Bastus, N. G.; Merkoci, F.; Piella, J.; Puntes, V. *Chemistry of Materials* **2014**, *26*, 2836–2846.

(S3) Dholakia, K.; Zemánek, P. *Reviews of Modern Physics* **2010**, *82*, 1767–1791.

(S4) Sukhov, S.; Shalin, A.; Haefner, D.; Dogariu, A. *Optics Express* **2015**, *23*, 247.

(S5) Figliozzi, P.; Sule, N.; Yan, Z.; Bao, Y.; Burov, S.; Gray, S. K.; Rice, S. A.; Vaikuntanathan, S.; Scherer, N. F. *Physical Review E* **2017**, *95*, 022604.

(S6) Golestanian, R.; Liverpool, T.; Ajdari, A. *New Journal of Physics* **2007**, *9*, 126.

(S7) Jiang, H.-R.; Yoshinaga, N.; Sano, M. *Physical Review Letters* **2010**, *105*, 268302.

(S8) Baffou, G.; Quidant, R.; García de Abajo, F. J. *ACS Nano* **2010**, *4*, 709–716.

(S9) Duhr, S.; Braun, D. *Proceedings of the National Academy of Sciences* **2006**, *103*, 19678–19682.

(S10) Piazza, R.; Parola, A. *Journal of Physics: Condensed Matter* **2008**, *20*, 153102.

(S11) Xu, Y.-l. *Applied Optics* **1995**, *34*, 4573–4588.

(S12) Ng, J.; Lin, Z.; Chan, C.; Sheng, P. *Physical Review B* **2005**, *72*, 085130.

(S13) Bohren, C. F.; Huffman, D. R. *Absorption and scattering of light by small particles*; John Wiley & Sons, 2008.

(S14) Birdsall, C. K.; Langdon, A. B. *Plasma physics via computer simulation*; CRC Press, 2004.

(S15) Sule, N.; Rice, S.; Gray, S.; Scherer, N. *Optics Express* **2015**, *23*, 29978–29992.

(S16) Sule, N.; Yifat, Y.; Gray, S. K.; Scherer, N. F. *Nano letters* **2017**, *17*, 6548–6556.

(S17) Shao, L.; Käll, M. *Advanced Functional Materials* **2018**, 1706272.